\documentclass[aps,prd,showpacs,floatfix,preprint]{revtex4}
\usepackage{amsmath,bm}
\usepackage{graphicx}
\usepackage{epsfig}
\begin{document}

\title{Bulk viscosity in kaon condensed matter}
\author{Debarati Chatterjee and Debades Bandyopadhyay} 
\affiliation{Saha Institute of Nuclear Physics, 1/AF Bidhannagar, 
Kolkata-700064, India}

\begin{abstract}
We investigate the effect of $K^-$ condensed matter on bulk viscosity and  
r-mode instability in neutron stars. The bulk viscosity coefficient due to the 
non-leptonic  process 
$n \rightleftharpoons p + K^-$ is studied here. In this connection, 
equations of state are constructed within the
framework of relativistic field theoretical models where nucleon-nucleon 
and kaon-nucleon interactions are mediated by the exchange of scalar and vector
mesons. We find that the bulk viscosity coefficient due to the non-leptonic 
weak process in the condensate is suppressed by several orders of magnitude. 
Consequently, kaon bulk viscosity may not damp
the r-mode instability in neutron stars. 
\end{abstract}
\maketitle

\section{Introduction}
Astrophysical compact objects such as neutron stars and black holes, are 
candidates of gravitational waves. Neutron stars pulsate in
different modes due to its fluid perturbation. The Coriolis restored inertial
r-modes of rapidly rotating neutron stars may be sources of detectable 
gravitational waves. Gravitational radiation 
drives the r-modes unstable due to Chandrasekhar-Friedman-Schutz mechanism 
\cite{Chan,Frie,Kok,And01,And98,Fri98,Lin98,And99,Ster}. On the other hand, 
this 
instability may be suppressed in different ways. One plausible explanation
for the damping of the instability is the large bulk viscosity coefficient due
to non-leptonic processes involving hyperons 
\cite{Jon1,Jon2,Lin02,Dal,Dra,Nar,Rati}. However, it may not be an efficient 
damping mechanism if hyperons are superfluid \cite{Nar,Han,And06}. Another 
possibility of damping the r-mode instability is the mutual friction between
inter-penetrating neutron and proton superfluids \cite{And06,Lin00}. This shows
that superfluidity of neutron star matter plays a significant role on the
damping of the r-modes in neutron stars.  

Besides hyperons, other novel phases with large strangeness fraction such as
quark matter and Bose-Einstein condensed matter of $K^-$ mesons, may appear in
neutron star interior. Already the bulk viscosity of unpaired quark matter and 
its influence on the r-mode instability were investigated by various groups 
\cite{Mad92,Mad00,Don1,Don2}.
Recently, bulk viscosity coefficients due to weak processes in quark-hadron 
mixed phase \cite{Dra,Pan} and color superconducting quark matter 
\cite{Bas,Schm,Alf,Bas2} have been studied by several groups. However, there 
are no calculations of bulk viscosity in Bose-Einstein condensed matter so far.
  
Kaplan and Nelson \cite{Kap} first demonstrated that Bose-Einstein condensation
of $K^-$mesons in dense hadronic matter was a possibility. This calculation was
performed using a chiral $SU(3)_L\times SU(3)_R$ Lagrangian. It was noted that 
the strongly attractive $K^-$-nucleon interaction lowered the effective mass and
in-medium energy of $K^-$ mesons in dense matter. The s-wave condensation sets 
in when the in-medium energy of negatively charged kaons equals to its chemical
potential. Kaon condensation in neutron star interior was investigated 
extensively in the chiral \cite{Bro92,Thor,Ell,Pra97} and 
the  meson exchange models \cite{Gle98,Gle99,Mut,Kno,Sch,Pon,Pal,Bani1,Bani2}. 
The onset of ${K^-}$ condensation in neutron star matter depends on the nuclear 
equation of state as well as on the depth of antikaon optical potential. Earlier
calculations showed that kaon condensation might set in at around two 
times normal nuclear matter density \cite{Bani2}. However, the early appearance
of hyperons may delay ${K^-}$ condensation to higher density or vice versa. 

With the onset of $K^{-}$ condensation, the neutron star matter is in chemical
equilibrium through the weak process $n \rightleftharpoons p + K^{-}$. 
The density perturbation associated with the r-modes drives the system out of
chemical equilibrium. The weak non-leptonic processes may reestablish the 
equilibrium. Therefore, it motivates us to investigate the bulk viscosity due 
to this non-leptonic process in kaon condensed matter and its influence on 
neutron star r-modes.  
The paper is organised in the following way. In section 2, we describe the 
field theoretical models of strong interactions, different phases of dense 
matter, the bulk viscosity coefficient and the corresponding time scale. 
Results of our calculation are explained in section 3. Section 4 gives
summary and conclusions. 

\section{Formalism}
Here we consider a first order phase transition from hadronic 
to $K^-$ condensed matter in neutron stars. Both 
the pure hadronic and $K^-$ condensed matter are described within the
framework of relativistic field theoretical models. The constituents 
of matter in both phases are neutrons ($n$), protons ($p$), electrons and muons.
Neutrons and protons embedded in the $K^-$ condensate behave dynamically 
differently than those of the hadronic phase. It was attributed to different 
mean fields which nucleons experienced in the pure phases 
\cite{Gle98,Gle99,Bani2}. 
Local charge neutrality and  beta-equilibrium conditions are satisfied in both
phases. The baryon-baryon interaction is mediated by the exchange
of scalar and vector mesons.  
Therefore the Lagrangian density for the pure hadronic phase is given by
\begin{eqnarray}
{\cal L}_B &=& \sum_{B=n,p} \bar\psi_{B}\left(i\gamma_\mu{\partial^\mu} - m_B
+ g_{\sigma B} \sigma - g_{\omega B} \gamma_\mu \omega^\mu
- g_{\rho B}
\gamma_\mu{\mbox{\boldmath t}}_B \cdot
{\mbox{\boldmath $\rho$}}^\mu \right)\psi_B\nonumber\\
&& + \frac{1}{2}\left( \partial_\mu \sigma\partial^\mu \sigma
- m_\sigma^2 \sigma^2\right) - U(\sigma) \nonumber\\
&& -\frac{1}{4} \omega_{\mu\nu}\omega^{\mu\nu}
+\frac{1}{2}m_\omega^2 \omega_\mu \omega^\mu
- \frac{1}{4}{\mbox {\boldmath $\rho$}}_{\mu\nu} \cdot
{\mbox {\boldmath $\rho$}}^{\mu\nu}
+ \frac{1}{2}m_\rho^2 {\mbox {\boldmath $\rho$}}_\mu \cdot
{\mbox {\boldmath $\rho$}}^\mu ~.
\end{eqnarray}
Here $\psi_B$ denotes the Dirac bispinor for baryons $B$ with vacuum mass $m_B$
and the isospin operator is ${\mbox {\boldmath t}}_B$. The scalar
self-interaction term \cite{Bog} is
\begin{equation}
U(\sigma) = \frac{1}{3} g_2 \sigma^3 + \frac{1}{4} g_3 \sigma^4 ~.
\end{equation}
We perform this calculation in the mean field approximation 
\cite{Ser}. The mean meson fields $\sigma$, $\omega_0$ and $\rho_{03}$
in the hadronic phase are given by
\begin{eqnarray}
m_\sigma^2\sigma &=& -\frac{\partial U}{\partial\sigma}
+ \sum_{B=n,p} g_{\sigma B} n_B^{{h},S}~,\\
m_\omega^2\omega_0 &=& \sum_{B=n,p} g_{\omega B} n_B^{h}~,\\
m_\rho^2\rho_{03} &=& \sum_{B=n,p} g_{\rho B} I_{3B} n_B^{h}~.
\end{eqnarray}
Here $n_B^{h,S}$ and $n_B^h$ are scalar and baryon number densities in the
hadronic phase. 

The total charge density of the hadronic phase is
\begin{equation}
Q^h = \sum_{B=n,p} q_B n^h_B -n_e -n_\mu =0~,
\end{equation}
where $q_B$ is the charge of baryons $B$ in 
the pure hadronic phase and $n_e$ and $n_\mu$ are charge densities of 
electrons and muons respectively. The $\beta$-equilibrium in this phase leads 
to $\mu_n^h = \mu_p^h + \mu_e$. 
The chemical potential for baryons $B$ is given by
\begin{equation}
\mu_{B}^{h} = \left(k^2_{F_{B}} + {m_B^{*h}}^2 \right)^{1/2} + g_{\omega B} 
\omega_0 + I_{3B} g_{\rho B} \rho_{03}~, 
\end{equation}
where effective baryon mass is $m_B^{*h}=m_B - g_{\sigma B}\sigma$ and
isospin projection for baryons $B$ is $I_{3B}$. 

The energy density and pressure in this phase are given by
\begin{eqnarray}
{\varepsilon^{h}}  &=& \frac{1}{2}m_\sigma^2 \sigma^2
+ \frac{1}{3} g_2 \sigma^3
+ \frac{1}{4} g_3 \sigma^4  
+ \frac{1}{2} m_\omega^2 \omega_0^2 
+ \frac{1}{2} m_\rho^2 \rho_{03}^2  \nonumber \\
&& + \sum_{B=n,p} \frac{2J_B+1}{2\pi^2}
\int_0^{k_{F_B}} (k^2+{m_B^{*h}}^2)^{1/2} k^2 \ dk
+ \sum_{l=e^-,\mu^-} \frac{1}{\pi^2} \int_0^{K_{F_l}} (k^2+m^2_l)^{1/2} k^2 \ dk
~,
\end{eqnarray}
\begin{eqnarray}
P^{h} &=& - \frac{1}{2}m_\sigma^2 \sigma^2 - \frac{1}{3} g_2 \sigma^3
- \frac{1}{4} g_3 \sigma^4  
+ \frac{1}{2} m_\omega^2 \omega_0^2 
+ \frac{1}{2} m_\rho^2 \rho_{03}^2 \nonumber \\
&& + \frac{1}{3}\sum_{B=n,p} \frac{2J_B+1}{2\pi^2}
\int_0^{k_{F_B}} \frac{k^4 \ dk}{(k^2+{m_B^{*h}}^2)^{1/2}}
+ \frac{1}{3} \sum_{l=e^-,\mu^-} \frac{1}{\pi^2}
\int_0^{K_{F_l}} \frac{k^4 \ dk}{(k^2+m^2_l)^{1/2}}~.
\end{eqnarray}

The baryon-baryon interaction in the pure kaon condensed phase has the same 
form as given by Eq. (1). In this case, we adopt a relativistic field
theoretical approach for the description of (anti)kaon-baryon interaction
\cite{Gle98,Gle99,Bani2}. Besides the exchange of $\sigma$, $\omega$ and
$\rho$ mesons, (anti)kaon-(anti)kaon interaction is also mediated by two 
strange mesons, scalar meson $f_0$(975) (denoted hereafter as $\sigma^*$) and 
the vector meson $\phi$(1020). As nucleons do not couple with strange mesons, 
$g_{\sigma^* N} = g_{\phi N} = 0$.
The Lagrangian density for (anti)kaons in the minimal coupling scheme
is, 
\begin{equation}
{\cal L}_K = D^*_\mu{\bar K} D^\mu K - m_K^{* 2} {\bar K} K ~,
\end{equation}
where the covariant derivative is
$D_\mu = \partial_\mu + ig_{\omega K}{\omega_\mu} + ig_{\phi K}{\phi_\mu}
+ i g_{\rho K}
{\mbox{\boldmath t}}_K \cdot {\mbox{\boldmath $\rho$}}_\mu$ and
the effective mass of (anti)kaons is 
$m_K^* = m_K - g_{\sigma K} \sigma - g_{\sigma^* K} \sigma^*$.
The mean meson fields in the condensed phase are denoted by $\sigma$, 
$\sigma^*$, $\omega_0$, $\phi_0$ and $\rho_{03}$. 
The in-medium energy of
$K^-$ mesons for $s$-wave (${\vec k}=0$) condensation is given by
\begin{equation}
\omega_{K^-} = \mu_{K^-} = m_K^* - g_{\omega K} \omega_0 - g_{\phi K} \phi_0
+ I_{3K^-} g_{\rho K} \rho_{03} ~,
\end{equation}
where $\mu_{K^-}$ is the chemical potential of $K^-$ mesons and the isospin 
projection is $I_{3K^-} = -1/2$. 
The mean fields in the $K^-$ condensed phase are
\begin{eqnarray}
m_\sigma^2\sigma &=& -\frac{\partial U}{\partial\sigma}
+ \sum_{B=n,p} g_{\sigma B} n_B^{{K},S}
+ g_{\sigma K} n_{K^-} ~,\\
{m_{\sigma^*}^2}\sigma^* &=& \sum_{B=n,p} g_{\sigma^* B} n_B^{{K},S}
+ g_{\sigma^* K} n_{K^-} ~,\\
m_\omega^2\omega_0 &=& \sum_{B=n,p} g_{\omega B} n_B^{K}
- g_{\omega K} n_{K^-} ~,\\
m_\phi^2\phi_0 &=& \sum_{B=n,p} g_{\phi B} n_B^{K}
- g_{\phi K} n_{K^-} ~,\\
m_\rho^2\rho_{03} &=& \sum_{B=n,p} g_{\rho B} I_{3B} n_B^{K}
+ g_{\rho K} I_{3K^-} n_{K^-} ~.
\end{eqnarray}
The scalar and baryon number densities in the condensed 
phase are given by
\begin{eqnarray}
n_B^{{K},S} &=& \frac{2J_B+1}{2\pi^2} \int_0^{k_{F_B}}
\frac{{m_B^{*K}}}{(k^2 + {m_B^{*K}}^ 2)^{1/2}} k^2 \ dk ~,\\
n_B^{K} &=& (2J_B+1)\frac{k^3_{F_B}}{6\pi^2} ~,
\end{eqnarray}
where Fermi momentum is $k_{F_B}$ and spin is $J_B$. 
The effective baryon mass ($m_B^{*K}$) in 
the condensate has the same form as in the hadronic phase. 
In the absence of source terms for $K^-$ mesons in Eq. (12)-(16), we 
retain the mean fields in the hadronic phase. 
The scalar density of $K^-$ mesons in the condensate is given 
by \cite{Gle99}
\begin{equation}
n_{K^-} = 2\left( \omega_{K^-} + g_{\omega K} \omega_0 
+ g_{\phi K} \phi_0 + \frac{1}{2} g_{\rho K} \rho_{03} \right) {\bar K} K  
= 2m^*_K {\bar K} K  ~.
\end{equation}

The total charge density in the condensed phase is 
\begin{equation}
Q^{K}=\sum_{B=n,p} q_B n_B^{K} -n_K^- - n_e - n_\mu =0
\end{equation}
With the onset of $K^-$ condensation, we have 
\begin{equation}
n \rightleftharpoons p + K^{-}~. 
\end{equation}
The requirement of chemical equilibrium yields
\begin{equation}
\mu_n^K - \mu_p^K = \mu_{K^-} = \mu_e ~.
\end{equation}
Here the baryon
chemical potential has the same form as given by Eq. (7), but the mean fields 
used here are obtained in the presence of the condensate. 

The total energy density in the kaon condensed phase is comprised of
baryons, leptons and antikaons
\begin{eqnarray}
{\varepsilon^{K}}  &=& \frac{1}{2}m_\sigma^2 \sigma^2
+ \frac{1}{3} g_2 \sigma^3
+ \frac{1}{4} g_3 \sigma^4  + \frac{1}{2}m_{\sigma^*}^2 \sigma^{*2}
+ \frac{1}{2} m_\omega^2 \omega_0^2 + \frac{1}{2} m_\phi^2 \phi_0^2
+ \frac{1}{2} m_\rho^2 \rho_{03}^2  \nonumber \\
&& + \sum_{B=n,p} \frac{2J_B+1}{2\pi^2}
\int_0^{k_{F_B}} (k^2+{m_B^{*K}}^2)^{1/2} k^2 \ dk
+ \sum_{l=e^-,\mu^-} \frac{1}{\pi^2} \int_0^{K_{F_l}} (k^2+m^2_l)^{1/2} k^2 \ dk
\nonumber \\
 && + m^*_K n_{K^-}~.
\end{eqnarray}
The last term denotes the contribution of the $K^-$ condensate. The pressure in this phase is
\begin{eqnarray}
P^{K} &=& - \frac{1}{2}m_\sigma^2 \sigma^2 - \frac{1}{3} g_2 \sigma^3
- \frac{1}{4} g_3 \sigma^4  - \frac{1}{2}m_{\sigma^*}^2 \sigma^{*2}
+ \frac{1}{2} m_\omega^2 \omega_0^2 + \frac{1}{2} m_\phi^2 \phi_0^2
+ \frac{1}{2} m_\rho^2 \rho_{03}^2 \nonumber \\
&& + \frac{1}{3}\sum_{B=n,p} \frac{2J_B+1}{2\pi^2}
\int_0^{k_{F_B}} \frac{k^4 \ dk}{(k^2+{m_B^{*K}}^ 2)^{1/2}}
+ \frac{1}{3} \sum_{l=e^-,\mu^-} \frac{1}{\pi^2}
\int_0^{K_{F_l}} \frac{k^4 \ dk}{(k^2+m^2_l)^{1/2}}~.
\end{eqnarray}

The mixed phase of hadronic and $K^-$ condensed matter is governed by  the 
Gibbs conditions for thermodynamic equilibrium along with global charge and
baryon number conservation laws 
\cite{Gle99,Gle92}. The Gibbs phase rules read,
\begin{eqnarray}
P^h&=& P^{K},\\
\mu_B^h& =& \mu_B^{K},
\end{eqnarray}
where $\mu_B^h$ and $\mu_B^{K}$ are chemical potentials of baryons B in the
pure hadronic and $K^-$ condensed phase, respectively.
The global charge neutrality and baryon number conservation laws are 
\begin{equation}
(1-\chi) Q^h + \chi Q^{K} = 0,\\
\end{equation}
\begin{equation}
n_B=(1-\chi) n_B^h + \chi n_B^{K}~,
\end{equation}
where $\chi$ is the volume fraction of $K^-$ condensed phase in the mixed 
phase. The total energy density in the mixed phase is given by 
\begin{equation}
\epsilon=(1-\chi)\epsilon^h + \chi \epsilon^{K}~.
\end{equation}

Pressure and density variations 
associated with the r-mode oscillation drive the system out of chemical
equilibrium. Microscopic reaction processes restore the 
equilibrium. Weak interaction processes are most important in this situation. 
The corresponding bulk viscosity coefficient could play a significant role on 
the mode. Earlier it was shown by various authors 
\cite{Nar,Jon1,Jon2,Lin02} that non-leptonic processes involving hyperons might
lead to a high value for the bulk viscosity coefficient. 
The real part of bulk viscosity coefficient is calculated in terms of 
relaxation times of microscopic processes \cite{Lin02,Lan} 
\begin{equation}
\zeta = \frac {P(\gamma_{\infty} - \gamma_0)\tau}{1 + {(\omega\tau)}^2}~,
\end{equation}
where $P$ is the pressure, $\tau$ is the net microscopic relaxation time and 
$\gamma_{\infty}$ and $\gamma_0$ are 'infinite' and 'zero' frequency adiabatic 
indices respectively. The factor
\begin{equation}
\gamma_{\infty} - \gamma_0 = - \frac {n_b^2}{P} \frac {\partial P} 
{\partial n_n} \frac {d{\bar x}_n} {dn_b}~,
\end{equation}
can be determined from the equation of state (EoS). Here 
$\bar x_n = \frac {n_n}{n_b}$ gives the 
neutron fraction in the equilibrium state and $n_b = {\sum}_{B} n_B$ is the 
total baryon density. In the co-rotating frame, the angular velocity 
($\omega$) of (l,m) r-mode is 
related to angular velocity ($\Omega$) of a rotating 
neutron star as $\omega = {\frac {2m}{l(l+1)}} \Omega$ \cite{And01}. 

The partial derivatives in both phases are calculated using the Gibbs-Duhem 
relation. In the pure hadronic phase, this relation gives
$P^h = n_n^h \mu_n^h + n_p^h \mu_p^h + n_e \mu_e + n_{\mu} \mu_e 
- \varepsilon^h$. So the partial 
derivative is given by 
\begin{equation}
\frac{\partial P^h}{\partial n_n^h} = \mu_n^h 
+ n_n^h \frac{\partial \mu_n^h}{\partial n_n^h} 
+ n_p^h \frac{\partial \mu_p^h}{\partial n_n^h} 
- \frac{\partial \varepsilon^h}{\partial n_n^h}~.
\end{equation}
Using $\mu_n^h = \frac{\partial \varepsilon^h}{\partial n_n^h}$, we obtain
\begin{equation}
\frac{\partial P^h}{\partial n_n^h} = n_n^h \alpha_{nn}^h 
+  n_p^h \alpha_{pn}^h~,
\end{equation} 
where $\alpha_{ij}$ is defined by
$\alpha_{ij} = \left(\frac{\partial \mu_i}{\partial n_j}\right)_{n_k,k\neq j}$. 
Similarly, in the pure condensed phase, the partial derivative is 
\begin{equation}
\frac{\partial P^K}{\partial n_n^K} = n_n^K \alpha_{nn}^K 
+  n_p^K \alpha_{pn}^K 
+ n_{K^-} \alpha_{K^- n}^K~.
\end{equation} 
In the mixed phase, we obtain the relation
\begin{equation}
\frac{\partial P}{\partial n_n} = 
\frac{\partial P^h}{\partial n_n^h} +  
\frac{\partial P^K}{\partial n_n^K}~, 
\end{equation} 
where $n_n = (1 - \chi) n_n^h + \chi n_n^K$.

We are interested to study the bulk viscosity coefficient and corresponding 
damping timescale due to the non-leptonic weak decay process (21). 
For the reaction rate of this process, we may express all perturbed quantities
in terms of the variation in neutron number density ($n_n^K$) in the 
Bose-Einstein condensed phase. The relaxation time ($\tau$) for the 
process is given by \cite{Lin02}
\begin{equation}
\frac{1}{\tau} = \frac{\Gamma_K}{\delta \mu} 
\frac{\delta \mu}{\delta n_n^K}~.
\end{equation}
Here, $\delta {n_n^K} = n_n^K - {\bar n}_n^K$ is the departure of neutron 
fraction from its thermodynamic equilibrium value ${\bar n}_n^K$ in the
$K^-$ condensed phase. 
The reaction rate per unit volume can be written as
\begin{equation}
{\Gamma_K} = \frac{{(2 \pi)}^4}{8{(2 \pi)}^9} 
\int \frac{d^3k_1}{E_1} \frac{d^3k_2}{E_2} \frac{d^3k_3}{E_3} 
{\langle |\mathcal{M}|^2 \rangle} \delta^3 (\vec k_1 - \vec k_2 - \vec k_3 ) 
F(E_i) \delta^0 (E_1 - E_2 - E_3)~.
\end{equation}
where indices $i=1-3$ refer to $n$, $p$ and $K^-$ respectively.
Here we are considering $K^-$ mesons in the $s$-wave ($\vec k_3=$0) condensate.
So the conservation of momentum gives 
$\vec k_1 = \vec k_2$.
Assuming that the matrix element has no angular dependence and performing the
integrating over $d^3 k_3$, we get
\begin{equation}
{\Gamma_K} = \frac{1}{8 {(2 \pi)}^5 E_3} 
\int \frac{|\vec k_1| E_1 dE_1} {E_1} \frac{|\vec k_2| E_2 dE_2}{E_2} 
{\langle |\mathcal{M}|^2 \rangle} F(E_i) \delta^0 (E_1 - E_2 - E_3) 
d\Omega_1 d\Omega_2~, 
\end{equation}
where $F(E_i)$ is the Pauli blocking factor given by
\begin{equation}
F(E_i) = f_1 (1-f_2) - (1-f_1)f_2~,
\end{equation} 
and $f_i = \frac{1}{(1+e^{\frac{E_i-\mu_i}{kT}})}$.
$|\mathcal{M}|^2$ is the squared matrix element and $\langle {.} \rangle$ 
denotes the averaging over initial spins and sum over final spins. In this
calculation fermions which reside close to their Fermi surfaces, contribute to
the reaction.
Integrating over $E_1$ and substituting $\frac{E_2-\mu_2}{kT}=y$ and 
$dE_2 = kT dy$, we obtain
\begin{equation}
{\Gamma_K}= \frac{\langle |\mathcal{M}|^2 \rangle}{16\pi^3E_3} k_{F_n}^2 
\left(\frac{\delta \mu}{k T}\right) kT \int_{ -\infty}^{+\infty} 
\frac{dy}{ (1+e^{-y}) (1+e^y)}~.
\end{equation} 
Finally we arrive at the result,
\begin{equation}
{\Gamma_K} = \left[ \frac{\langle |\mathcal{M}|^2 \rangle}{16\pi^3E_3} 
k_{F_n}^2 \right] \delta \mu~.
\end{equation} 
Here the in-medium energy of $K^-$ mesons in the condensate is
$E_3 = \mu_3 = \mu_{K^-}$ and $k_{F_n}$ is the Fermi momentum for neutrons in 
the condensed phase. 
It is noted that the reaction rate per unit volume is 
proportional to the overall chemical potential imbalance 
$\delta \mu = \delta \mu_n^K - \delta \mu_p^K - \delta \mu_{K^-}$. 
Now the relaxation time (36) can be written as
\begin{equation}
\frac{1}{\tau} =  \left[ \frac{\langle |\mathcal{M}|^2 \rangle}
{16\pi^3 \mu_{K^-}} 
k_{F_n}^2 \right] \frac{\delta \mu}{\delta n_n^K}~.
\end{equation}

Next we focus on the evaluation of the matrix element of the non-leptonic 
process (21). In the general form, this can be written as \cite{Com,Oku} 
\begin{equation}
\mathcal{M} = \bar u(k_2) \left( A + B \gamma_5 \right) u(k_1)~,
\end{equation}
where  $u(k_1)$ and $u(k_2)$ are the spinors of neutrons and 
protons respectively. Here A is the parity violating amplitude and B is the 
parity conserving amplitude. 
The squared matrix element is given by
\begin{equation}
{\langle |\mathcal{M}|^2 \rangle} =  { 2 [\left( k_1 \cdot k_2 + m_n^{*K} 
m_p^{*K} 
\right) |A|^2 + \left( k_1 \cdot k_2 - m_n^{*K} m_p^{*K} \right) |B|^2]}~.
\end{equation}
As fermion momenta lie close to their Fermi surfaces, we can write  
$k_1 \cdot k_2 = E_1 E_2 - |\vec k_1||\vec k_2| \cos \theta 
= \mu_n^K \mu_p^K - k_{F_n}^2$. This leads to the squared matrix element
\begin{equation}
{\langle |\mathcal{M}^2| \rangle} = 2  [\left(\mu_n^K \mu_p^K - k_{F_n} k_{F_p} 
+ m_n^{*K} m_p^{*K} \right)|A|^2 
+ \left(\mu_n^K \mu_p^K - k_{F_n} k_{F_p} - m_n^{*K} m_p^{*K} \right)|B|^2 ]~.
\end{equation}

Here we apply the weak SU(3) symmetry to the non-leptonic weak decay amplitudes
for the process (21). Earlier weak decays of the octet hyperons were
described by an effective SU(3) interaction with a parity violating (A) and 
parity conserving (B) amplitudes \cite{Mar,Sch00}. The weak operator is 
proportional to Gell-Mann matrix $\lambda_6$ to ensure hypercharge
violation $|\Delta Y| = 1$ and $|\Delta I| = 1/2$ rule. The amplitudes were
determined experimentally from the weak decay parameters of the octet hyperons
\cite{Sch00}. We follow the same mechanism to determine the amplitudes
in (21). The amplitudes are $A = - 1.62 \times 10^{-7}$ and 
$B = -7.1 \times 10^{-7}$.

The computation of the relaxation time is complete with the calculation of 
$\frac {\delta \mu}{\delta{n_n^K}}$. The chemical potential imbalance due to 
the non-leptonic process (21) is given by
\begin{eqnarray}
\delta \mu &=& \delta \mu_n^K - \delta \mu_p^K - \delta \mu_{K^-} \nonumber \\
&=& \left( \alpha_{nn}^K \delta n_n^K + \alpha_{np}^K \delta n_p^K + 
\alpha_{n K^-} \delta n_{K^-} \right) \nonumber \\
&-& \left( \alpha_{pn}^K \delta n_n^K + \alpha_{pp}^K \delta n_p^K + 
\alpha_{p K^-} \delta n_{K^-} \right) \nonumber \\
&-& \left( \alpha_{K^- n} \delta n_n^K + \alpha_{K^- p} \delta n_p^K + 
\alpha_{K^- K^-} \delta n_{K^-} \right)~.
\end{eqnarray}
We express $\delta {\mu}$ in terms of $\delta{n_n^K}$ and obtain 
$\frac {\delta \mu}{\delta{n_n^K}}$ from the following constraints,
\begin{eqnarray}
(1-\chi)(\delta n_n^h + \delta n_p^h) + \chi (\delta n_n^K + \delta n_p^K) 
&=& 0~, \nonumber \\
(1-\chi)\delta n_p^h + \chi ( \delta n_p^K - \delta n_{K^-}) &=& 0~, 
\nonumber \\
\delta \mu_p^h &=& \delta \mu_p^K~, \nonumber \\
\delta \mu_n^h &=& \delta \mu_n^K~. 
\end{eqnarray} 
In the above constraints, we have $\delta \chi = 0$ because number densities
deviate from their equilibrium values not by changing the volume of the fluid 
element but by internal reactions \cite{Lin02}.
First two constraints result from the conservation of baryon number and
electric charge neutrality. The other constraints are due to the equality of
neutron and proton chemical potentials in the hadronic and condensed phases.
Further the last two constraints in (47) can be written as 
\begin{eqnarray}
\left(\alpha_{pn}^h \delta n_n^h + \alpha_{pp}^h \delta n_p^h \right) -   
\left(\alpha_{pn}^K \delta n_n^K + \alpha_{pp}^K \delta n_p^K +   
\alpha_{p K^-} \delta n_{K^-}\right) &=& 0~, \nonumber \\ 
\left(\alpha_{nn}^h \delta n_n^h + \alpha_{np}^h \delta n_p^h \right) -   
\left(\alpha_{nn}^K \delta n_n^K + \alpha_{np}^K \delta n_p^K +   
\alpha_{n K^-} \delta n_{K^-} \right) &=& 0~. 
\end{eqnarray} 

Next we calculate $\alpha_{ij}$ in the hadronic as well as $K^-$ condensed
phases using the EoS. These quantities in the hadronic phase are given by 
\begin{eqnarray}
\alpha_{nn}^h &=& \frac{\partial \mu_n^h}{\partial n_n^h} \nonumber \\
&=& \left(\frac{g_{\omega N}}{m_{\omega}}\right)^2 
+ \frac{1}{4}\left(\frac{g_{\rho N}}{m_{\rho}}\right)^2 
+ \frac{\pi^2}{k_{F_n} \sqrt{k_{F_n}^2 
+ {m_N^{*h}}^2}} - \frac{m_N^{*h}g_{\sigma N}}{\sqrt{k_{F_n}^2 
+ {m_N^{*h}}^2}} \frac{\partial \sigma}{\partial n_n^h}~,
\end{eqnarray} 
\begin{equation}
\alpha_{np}^h = \left(\frac{g_{\omega N}}{m_{\omega}}\right)^2 
- \frac{1}{4}\left(\frac{g_{\rho N}}{m_{\rho}}\right)^2 
- \frac{m_N^{*h}g_{\sigma N}}{\sqrt{k_{F_n}^2 
+ {m_N^{*h}}^2}} \frac{\partial \sigma}{\partial n_p^h}~,
\end{equation} 
where
\begin{equation}
\frac{\partial \sigma}{\partial n_n^h} = 
\frac{\left(\frac{g_{\sigma N}}{m_{\sigma}^2}\right)
\frac{m_N^{*h}}{\sqrt{k_{F_n}^2 + {m_N^{*h}}^2}}}{D}~,
\end{equation}
\begin{equation}
\frac{\partial \sigma}{\partial n_p^h} = 
\frac{\left(\frac{g_{\sigma N}}{m_{\sigma}^2}\right)
\frac{m_N^{*h}}{\sqrt{k_{F_p}^2 + {m_N^{*h}}^2}}}{D}~,
\end{equation} 
and
\begin{equation}
D = { 1+\frac{1}{m_\sigma^2}\frac{d^2U}{d \sigma^2} 
+ \sum_{B=n,p} \frac{(2J_B+1)}{2 \pi^2}\left(\frac{g_{\sigma B}}{m_\sigma}
\right)^2 \int_0^{K_{F_B}} \frac{k^4 dk}{(k^2 + {m_B^{*h}}^2)^{3/2}}}~.
\end{equation} 
Also, $\alpha_{pp}^h$ and  $\alpha_{pn}^h$ are obtained by interchanging 
$n \leftrightarrow p$ in $\alpha_{nn}^h$ and $\alpha_{np}^h$ respectively. 
Similarly, $\alpha_{ij}$ in the $K^-$ condensed phase are given by
\begin{equation}
\alpha_{nn}^K = \frac{\partial \mu_n^K}{\partial n_n^K}\\
= \left(\frac{g_{\omega N}}{m_{\omega}}\right)^2 
+ \frac{1}{4}\left(\frac{g_{\rho N}}{m_{\rho}}\right)^2 
+ \frac{\pi^2}{k_{F_n} \sqrt{k_{F_n}^2 
+ {m_N^{*K}}^2}} - \frac{m_N^{*K}g_{\sigma N}}{\sqrt{k_{F_n}^2 
+ {m_N^{*K}}^2}} \frac{\partial \sigma}{\partial n_n^K}~,
\end{equation}
\begin{equation}
\alpha_{np}^K = \left(\frac{g_{\omega N}}{m_{\omega}}\right)^2 
- \frac{1}{4}\left(\frac{g_{\rho N}}{m_{\rho}}\right)^2 
- \frac{m_N^{*K}g_{\sigma N}}{\sqrt{k_{F_n}^2 
+ {m_N^{*K}}^2}} \frac{\partial \sigma}{\partial n_p^K}~,
\end{equation}
\begin{equation}
\alpha_{n K^-} = \frac{\partial \mu_n^K}{\partial n_{K^-}} \\
= -\left( \frac{g_{\omega N}g_{\omega K}}{m_\omega^2}\right) 
+ \frac{1}{4}\left( \frac{g_{\rho N}g_{\rho K}}{m_\rho^2}\right) 
- \frac{m_N^{*K}g_{\sigma N}}{\sqrt{k_{F_n}^2 
+ {m_N^{*K}}^2}} \frac{\partial \sigma}{\partial n_{K^-}}~,
\end{equation}
\begin{equation}
\alpha_{K^- n} = \frac{\partial \mu_{K^-}}{\partial n_n^K} \\
= -\left( \frac{g_{\omega N}g_{\omega K}}{m_\omega^2}\right) 
+ \frac{1}{4}\left( \frac{g_{\rho N}g_{\rho K}}{m_\rho^2}\right) 
- g_{\sigma K} \frac{\partial \sigma}{\partial n_n^K}~,
\end{equation}
\begin{equation}
\alpha_{K^- K^-} = \frac{\partial \mu_{K^-}}{\partial n_{K^-}} \\
= \left( \frac{g_{\omega K}}{m_{\omega}} \right)^2 
+ \left( \frac{g_{\phi K}}{m_{\phi}} \right)^2 
+ \frac{1}{4}\left( \frac{g_{\rho K}}{m_{\rho}} \right)^2 
- g_{\sigma K} \frac{\partial \sigma}{\partial n_{K^-}} 
- g_{\sigma^* K} \frac{\partial \sigma^{*}}{\partial n_{K^-}}~,
\end{equation}
where,
\begin{eqnarray}
\frac{\partial \sigma}{\partial n_n^K} &=& \left( \frac{g_{\sigma N}}
{m_{\sigma}^2} \right) 
\frac{ \frac{{ m_N^{*K}}}{{ \sqrt{k_{F_n}^2 + {m_N^{*K}}^2 }} } }{D'},
\nonumber \\
\frac{\partial \sigma}{\partial n_{K^-}} &=& \frac{\frac{g_{\sigma K}}
{m_{\sigma}^2}}{D'},\nonumber\\
\frac{\partial \sigma^{*}}{\partial n_{K^-}} &=& \frac{g_{\sigma^* K}}
{m_{\sigma^*}^2},
\end{eqnarray} 
and $\frac{\partial \sigma^{*}}{\partial n_n^K} = 
\frac{\partial \sigma^{*}}{\partial n_p^K} = 0$. Here $D'$ has the same form
as $D$ in (53), but $D'$ has to be calculated using the meson fields in the
condensed phase.
The coefficients $\alpha_{pp}^K$ , $\alpha_{pn}^K$ , $\alpha_{p K^-}$ and 
$\alpha_{K^- p}$ are obtained by interchanging $n \leftrightarrow p$ in 
$\alpha_{nn}^K$ , $\alpha_{np}^h$ , $\alpha_{n K^-}$ and $\alpha_{K^- n}$ 
respectively.

The bulk viscosity damping timescale ($\tau_B$) due to the 
non-leptonic process (21) is defined by 
\begin{equation}
{\frac {1}{\tau_B}} =  - {\frac {1} {2E}} {\frac {dE}{dt}}~,
\end{equation}
where E is the energy of the perturbation in the co-rotating frame
of the fluid and is given by \cite{Lin02}
\begin{equation}
E = \frac {1}{2}{\alpha^2}{\Omega^2}{R^{-2}} \int_0^R {\epsilon (r) r^6}dr.
\end{equation}
Here $\alpha$ is the dimensionless amplitude of the r-mode, R is the 
radius of the star and $\epsilon$(r) is the energy density profile. The 
rate of energy dissipation is given by
\begin{equation}
\frac {dE}{dt} = -4 \pi \int_0^R \zeta (r) <|\vec{\nabla} \cdot 
{\delta \vec{v}}|^2> r^2 dr,
\end{equation} 
where the angle average of the square of the hydrodynamic expansion 
\cite{Nar,Lin99} is $\langle |\vec{\nabla}\cdot{\delta \vec{v}}|^2 \rangle 
= \frac {({\alpha \Omega})^2}{690}
\left({\frac {r}{R}}\right)^6 \left[1 
+ 0.86 \left({\frac {r}{R}}\right)^2\right] \left({\frac {\Omega^2}
{\pi G \bar {\epsilon}}}\right)^2$
and $\bar {\epsilon}$ is the mean energy density of a non-rotating star. 
Equation (60) is the bulk viscosity contribution to the imaginary 
part of the r-mode frequency. We also take into account time scales associated 
with gravitational radiation ($\tau_{GR}$), bulk viscosity due to modified Urca
process ($\tau_U$) involving only nucleons and the shear viscosity 
($\tau_{SV}$) and define the overall r-mode time scale ($\tau_r$) as
\begin{equation}
{\frac {1}{\tau_r}} =  - {\frac {1}{\tau_{GR}}} + {\frac {1}{\tau_B}} + 
{\frac {1}{\tau_U}} + {\frac {1}{\tau_{SV}}}~. 
\end{equation}
The gravitational radiation timescale is given by \cite{Lin98} 
\begin{equation}
{\frac {1}{\tau_{GR}}} =  
\frac {131072 \pi}{164025} {\Omega^6} \int_0^R {\epsilon (r) r^6}dr~.
\end{equation}
The time scale corresponding to the bulk viscosity due to the modified Urca 
process ($\tau_U$) involving only nucleons is calculated in the same footing 
as the bulk viscosity damping time scale due to the 
non-leptonic process but using the following expression 
for the bulk viscosity coefficient for the modified Urca process given by 
\cite{Kok,Saw}
\begin{equation}
\zeta_U = 6 \times 10^{25} \left({\frac {\epsilon}{10^{15} g/cm^3}}\right)^2 
\left({\frac{T}{10^9 K}}\right)^6 \left({\frac{\omega}{1 Hz}}\right)^{-2} 
g/cm s~.
\end{equation}
The damping time scale due to the shear viscosity is given by \cite{Lin98}
\begin{equation}
{\frac {1}{\tau_{SV}}} =  5 \int_0^R {\eta (r) r^4}dr
\left(\int_0^R {\epsilon (r) r^6}dr \right)^{-1}~.
\end{equation}
Different scattering processes contribute to the shear viscosity ($\eta$)
depending on whether neutrons and protons are superfluid or not 
\cite{Kok,And06}. In this calculation, neutrons and protons are both normal 
fluids. The most important contribution comes from neutron-neutron scattering 
in this case and the shear viscosity coefficient has the form \cite{Kok}, 
\begin{equation}
\eta = 2 \times 10^{18} \left({\frac{\epsilon}{10^{15} g/cm^3}}\right)^{9/4} 
\left({\frac{T}{10^9 K}}\right)^{-2} g/cm s~.
\end{equation}

The total r-mode time scale is a function of angular velocity, neutron star mass
and temperature. 
Therefore, solving $\frac {1}{\tau_r}$ = 0, we calculate the critical 
angular velocity above which the r-mode is unstable whereas it is stable below 
the critical angular velocity. 
\section{Results and Discussion}
We need to know meson-nucleon and meson-kaon coupling constants for this
calculation. The nucleon-meson coupling constants are determined from the
nuclear matter saturation
properties of binding energy $E/B=-16.3$ MeV, baryon density $n_0=0.153$ 
fm$^{-3}$, asymmetry energy coefficient $a_{\rm asy}=32.5$ MeV, 
incompressibility $K=240$ MeV and effective nucleon mass $m^*_N/m_N = 0.70$. 
The coupling constants are taken from Ref.\cite{Gle91} and 
this set is known as GM.  
We determine kaon-meson coupling constants using
the quark model and isospin counting rule. The vector coupling constants are
given by
\begin{equation}
g_{\omega K} = \frac{1}{3} g_{\omega N} ~~~~~ {\rm and} ~~~~~
g_{\rho K} = g_{\rho N} ~.
\end{equation}
The scalar coupling constant is obtained from the real part of
$K^-$ optical potential depth at normal nuclear matter density
\begin{equation}
U_{\bar K} \left(n_0\right) = - g_{\sigma K}\sigma - g_{\omega K}\omega_0 ~.
\end{equation}
It was found in various calculations that antikaons experienced an 
attractive potential and kaons had a repulsive interaction in nuclear matter 
\cite{Fri94,Fri99,Koc,Waa,Li,Pal2}. The strength of antikaon optical potential
ranges from shallow ($-40$ MeV) to strongly attractive ($-180$ MeV).
Moreover, it was noted that $K^-$ condensation could be a possibility for 
antikaon potential $\sim$ 100 MeV or more attractive. Here we perform the
calculation for a set of values of antikaon optical potential depths at normal 
nuclear matter density. We obtain kaon-scalar meson coupling 
constants $g_{\sigma K} = 2.1914, 3.0426, 3.8937$ corresponding to 
$U_{\bar K}(n_0) = -100, -120, -140$ MeV. 

The strange meson fields couple with (anti)kaons.
The $\sigma^*$-K coupling constant is $g_{\sigma^*K}=2.65$ as 
determined from the decay of $f_0$(925) and the vector $\phi$ meson
coupling with (anti)kaons $\sqrt{2} g_{\phi K} = 6.04$ follows from
the SU(3) relation \cite{Sch}. 

We begin our investigation with $K^-$ condensation in nuclear matter and denote
this system as np$K^-$. Baryon densities corresponding to lower ($u_l=n_l/n_0$)
and
upper ($u_u=n_u/n_0$) boundaries for $K^-$ condensation with antikaon optical 
potential depths at normal nuclear matter density $U_{\bar K} (n_0) = -100, 
-120, -140$ MeV are shown in Table I. The $K^-$ condensation sets in at a 
density $n_l=u_l n_0$ and a pure $K^-$ condensed phase with nucleons embedded 
in it begins at a density $n_u=u_u n_0$. It is evident from the Table that 
there is no 
mixed phase for $U_{\bar K}(n_0) = - 100$ MeV. We further note that the phase 
transition is of second order for $|U_{\bar K}(n_0)| \leq 100$ MeV. 
On the other hand, the onset of $K^-$ condensation shifts towards lower 
density as the antikaon optical potential becomes more attractive. The mixed 
phase is found to be wider for $U_{\bar K}(n_0) = -140$ MeV. These findings 
were already  reported in Ref. \cite{Gle99}. Another interesting problem arises
when hyperons appear around threshold densities of $K^-$ condensation. We 
showed in an earlier calculation \cite{Rati} by extending the Lagrangian 
density in (1) to include hyperons that $\Lambda$ hyperons first appeared 
around baryon density 2.58$n_0$. However, $\Sigma$ hyperons are excluded from
the system because of a strongly repulsive $\Sigma$-nuclear matter 
interaction whereas massive $\Xi$ hyperons appear at higher densities. In this 
context, we continue our calculation with $K^-$ condensation in np$\Lambda$ 
matter denoted by np${\Lambda} K^-$. In this case, $\Lambda$-meson couplings are
taken from Ref.\cite{Rati}. The threshold density ($u_{th}^{\Lambda}$) for the
appearance of $\Lambda$ hyperons are listed in Table I. For 
$U_{\bar K}(n_0) = -100, -120$ MeV, $\Lambda$ hyperons appear before the $K^-$ 
condensate. Consequently, $K^-$ condensation is delayed to higher densities.
Those values are recorded in parentheses. On the other hand, the early 
occurrence of $K^-$ condensation for $U_{\bar K}(n_0) = -140$ MeV delays the 
appearance of $\Lambda$ hyperons to higher density. Therefore, the composition 
of matter and the strength of antikaon optical potential depth determine the 
onset of $K^-$ condensate and $\Lambda$ hyperons in np${\Lambda} K^-$ matter. 
It was already noted that the bulk viscosity coefficient due to the 
non-leptonic process involving $\Lambda$ hyperons had a very large value. In 
this calculation, we study the bulk viscosity due to non-leptonic process
(21) in np$K^-$ matter and find out whether kaon bulk viscosity alone can damp
the r-mode without hyperons.

Now we describe the composition of $\beta$-equilibrated and charge neutral 
matter in the presence of $K^-$ condensation for $U_{\bar K}(n_0) = -120$ MeV. 
Populations of various particles
are displayed in Fig. 1. Before the appearance of $K^-$ condensate, particle
fractions increase with baryon density. In this situation, charge neutrality is
maintained among protons, electrons and muons. The mixed phase begins with the 
onset of $K^-$ condensation at 3.26 $n_0$. As soon as $K^-$ condensate is 
formed, its rapid growth replaces electrons and muons from the system. 
Consequently, the proton density becomes equal to the density of $K^-$ mesons 
in the condensate.    

The equations of state, pressure versus energy density, with and without 
$K^-$ condensate are shown in Fig. 2. The solid
line represents the overall EoS without $K^-$ condensate whereas the long 
dashed, short dashed and dotted lines correspond to those with the condensate 
for $U_{\bar K} = -100, -120, -140$ MeV respectively. We have already noted 
that the phase transition is of second order for $U_{\bar K}(n_0) = -100$ MeV. 
On the other hand, we find that two kinks on the EoS involving the condensate 
mark the 
beginning and end of the mixed phase for $U_{\bar K}(n_0) = -120, -140$ MeV. 
The EoS becomes softer in the presence of $K^-$ condensate compared with the 
case without the condensate. The deeper the antikaon optical potential depth, 
the softer is the EoS. 

Next we construct models of non-rotating and uniformly rotating neutron stars
\cite{Ster95} using EoS as shown in Fig. 2. The gravitational mass for 
non-rotating neutron star sequence and the sequence of neutron stars rotating 
at their Kepler frequencies are plotted with central energy density 
$\epsilon_c$ in Fig. 3. The inclusion of $K^-$ condensate in EoS for different 
values of $U_{\bar K}(n_0)$ reduces maximum masses for non-rotating as well as 
rotating neutron stars. The maximum gravitational masses, equatorial radii, 
corresponding central baryon and energy densities and Kepler periods for the 
mass shedding limit are given by Table II. The corresponding values for 
non-rotating neutron stars are shown in parentheses. It is evident from 
Table II that the softest EoS corresponding to $U_{\bar K}(n_0) = -140$ MeV 
leads to a maximum non-rotating neutron star mass 1.343 $M_{\odot}$ 
which is well below the most accurately measured Hulse-Taylor pulsar
mass 1.44 $M_{\odot}$. Therefore, this EoS is ruled out by observations. 
We adopt antikaon optical potential depths $U_{\bar K}(n_0) = -100, -120$ 
MeV for the calculation of bulk viscosity.   

The equation of state enters as an input in the calculations of adiabatic 
indices, relaxation time and bulk viscosity coefficient. In particular, partial
derivatives
of pressure and chemical potentials with respect to neutron number density are
calculated using the EoS. Figure 4 exhibits the difference of fast and slow
adiabatic indices as a function of normalised baryon density for 
$U_{\bar K}(n_0) = -100, -120$ MeV. We find jumps in the difference of adiabatic
indices ($\gamma_{\infty} - \gamma_0$). For $U_{\bar K}(n_0) = -100$ MeV, this
jump occurs at the onset of $K^-$ condensation whereas these are
attributed to kinks on the EoS for $U_{\bar K} = -120$ MeV in Fig. 2. Those
kinks lead to discontinuities in the partial derivatives in Eq. (31). 
This kind of jump in adiabatic indices was earlier noted in quark-hadron phase 
transition in neutron star matter by others \cite{Gen}. We further note that 
the value of ($\gamma_{\infty} - \gamma_0$) is suppressed for 
$U_{\bar K}(n_0) = -100$ MeV compared with the case of $U_{\bar K}(n_0) = -120$
MeV. The relaxation time for the 
non-leptonic weak process (21) is plotted with normalised baryon density in
Fig. 5 for different values of antikaon optical potential depths. In this case,
the maximum value of relaxation time is $\sim 10^{-11}s$ for 
$U_{\bar K}(n_0) = -120$ MeV. 
This value is a few order of magnitude smaller than that of the non-leptonic 
process involving non-superfluid $\Lambda$ hyperons \cite{Nar,Rati}. 
Also, we note that the
relaxation time given by Eq. (42) is independent of temperature because there
is no temperature dependence in the reaction rate in (41). This may be 
attributed to the fact that the $K^-$ condensate is treated here as a zero 
temperature Bose-Einstein condensate. On the other hand, the relaxation time
for the non-leptonic weak process involving hyperons was found to be inversely 
proportional to $T^2$ \cite{Lin02,Nar,Rati}. It is worth mentioning here that 
the role of thermal kaon processes in the non-equilibrium dynamics of kaon 
condensation was studied by others \cite{Muto1,Muto2}. 

The bulk viscosity coefficient in the kaon condensed phase is shown as a
function of normalised baryon density in Fig. 6. The factor $\omega \tau$ in 
the bulk viscosity coefficient given by Eq. (30) is negligible compared with 
unity over the whole range of baryon densities considered here. We note that 
the bulk viscosity coefficient is discontinuous at the phase boundaries of
hadronic and $K^-$ condensed matter. This is attributed to kinks in the EoS
and discontinuities in ($\gamma_{\infty} - \gamma_0$).
It is worth mentioning here that the
bulk viscosity coefficient in the kaon condensed matter is several orders
of magnitude smaller than the non-superfluid hyperon bulk viscosity coefficient 
\cite{Lin02,Nar,Rati}. We also note that the bulk viscosity coefficient is 
larger for $U_{\bar K}(n_0) = -120$ MeV than that of the case with 
$U_{\bar K}(n_0) = -100$ MeV in the lower density regime. We use
the bulk viscosity coefficient corresponding to $U_{\bar K}(n_0) = -120$ MeV 
for the calculation of damping time scale and critical angular velocity. 
 
The computation of damping time scale due to bulk viscosity given by Eq. (60) 
involves the energy density profile, kaon bulk viscosity profile and structures 
of rotating neutron stars. The sequence of neutron stars rotating at Kepler 
frequencies is calculated by using the model of Stergioulas \cite{Ster95}. We 
have 
already shown the mass shedding limit sequences of rotating neutron stars in 
Fig. 3. For the calculation of damping time scale, rotating neutron stars which
contain $K^-$ condensate in its interior are to be considered. The appearance 
of $K^-$ condensate is sensitive to the central density of a rotating neutron
star which ,in turn, depends on the rotation period of the star. It is 
worth mentioning here that the canonical neutron star of 1.4$M_{\odot}$ rotating
at the Kepler period 1.22 ms has a central baryon density 2.20$n_0$ and that of
its non-rotating counterpart has a central baryon density 2.76$n_0$ for 
$U_{\bar K}(n_0) = -100, -120$ MeV. This shows that central densities of 
1.4$M_{\odot}$ canonical neutron star are well below threshold densities of
$K^-$ condensation. To demonstrate the effect of kaon bulk viscosity on the 
r-mode instability,  we choose a neutron star of gravitational mass 
1.63$M_{\odot}$ having central baryon density 3.94 $n_0$ and 
rotating at an angular velocity $\Omega_{rot} = 1180 s^{-1}$ from the sequence
of rotating neutron stars corresponding to $U_{\bar K} = -120$ MeV. This 
neutron star contains $K^-$ condensate in its core because the central baryon 
density is well above the threshold density of $K^-$ condensation. 
The energy density profile of this rotating 
star is plotted as a function of equatorial distance in Fig. 7. Similarly, the
kaon bulk viscosity profile of this neutron star as a function of equatorial 
distance is shown in Fig. 8.
Here we note that the bulk viscosity profile drops to zero value beyond 2 km.
This happens because the baryon density beyond 2 km decreases below the 
threshold density of $K^-$ condensation and the non-leptonic process
(21) ceases to occur there.

Using the energy density and bulk viscosity 
profiles, we estimate the damping time scale corresponding to the bulk 
viscosity due to the non-leptonic process (21). We also consider the 
contributions of modified Urca process involving nucleons as well as the shear
viscosity to the total r-mode time scale as given by Eq. (63). We obtain 
critical angular velocities as a function of temperature solving 
$1/\tau_r = 0$ for a rotating neutron star mass 1.63$M_{\odot}$ corresponding
to the case $U_{\bar K}(n_0) = -120$ MeV and show 
this in Fig. 9. The r-mode is unstable above the critical angular velocity 
curve and stable below it. The bulk viscosity due to the modified Urca process 
dominates at higher temperatures and might damp the r-mode instability. 
The long dashed line which merges with the solid line above 5$\times 10^{9}$K,
shows the modified Urca contribution balancing the gravitational radiation 
reaction. As temperature decreases, the bulk viscosity due to the modified Urca
process also decreases. Consequently the corresponding damping time scale due 
to the modified Urca process is longer than the gravitational radiation growth
time scale and the r-mode instability is not suppressed below $10^{9}$ K by 
this process. Now we consider the shear viscosity. As the shear viscosity
coefficient is proportional to $T^{-2}$, it might suppress the instability at 
lower temperatures. This is shown by the short dashed line which merges with 
the solid line below temperature 5$\times 10^{9}$K. Next we discuss what is the 
influence of kaon bulk viscosity on the r-mode instability. We know already 
that kaon bulk viscosity is independent of temperature. The bulk 
viscosity damping time scale ($\tau_B$) due to the non-leptonic process (21) 
and the time scale associated with gravitational radiation are also independent
of temperature but depend on angular velocity ($\Omega$). Further investigation
reveals that the damping time scale $\tau_B$ is always longer than 
the gravitational radiation growth time scale 
$\tau_{GR}$ for all values of $\Omega$ ranging from 0 to $\Omega_{rot}$.
This implies that the bulk viscosity due to the non-leptonic process in $K^-$ 
condensed matter can not damp the r-mode instability. When all viscous 
processes 
are taken into account, we find an intermediate temperature regime where 
neither shear viscosity nor bulk viscosity can damp the r-mode instability, is
marked by the solid line. The critical angular velocity curve is determined due
to the interplay of $\tau_U$, $\tau_{SV}$ and $\tau_{GR}$. 

\section{Summary and Conclusions}       
We have investigated the role of $K^-$ condensation on bulk viscosity and 
r-mode instability in neutron stars. The bulk viscosity coefficient and the 
corresponding damping time
scale due to the non-leptonic process $n \rightleftharpoons p + K^{-}$ 
have been calculated. In this calculation, we have considered first and second
order kaon condensation and constructed equations of state for different values
of antikaon optical potential depths within the framework of relativistic field
theoretical models. It is noted that kaon bulk
viscosity and the corresponding damping time scale are independent of 
temperature. We find that the bulk
viscosity coefficient in $K^-$ condensed matter is suppressed by several orders
of magnitude compared with the non-superfluid hyperon bulk viscosity 
coefficient\cite{Lin02,Nar,Rati}. Further we note 
that kaon bulk viscosity due to the process (21) alone can not damp the r-mode
instability because kaon bulk viscosity damping time scale is always longer 
than the gravitational radiation growth time scale. 

We have also discussed the role of hyperons on the appearance of $K^-$ 
condensate. The early appearance of $\Lambda$ hyperons delays the onset of
$K^-$ condensation to higher densities for antikaon optical potential depths
$U_{\bar K}(n_0) = -100, -120$ MeV. For $U_{\bar K}(n_0) = -140$ MeV, $K^-$ 
condensation occurs before $\Lambda$ hyperons are populated in the system.
The latter situation was already reported in Ref. \cite{Bani2} for a different
parameter set. In this connection, we note that the large value of hyperon bulk 
viscosity might damp the r-modes in neutron stars when hyperons appear in
neutron stars. On the other hand, the r-modes in neutron stars 
containing only kaon condensate could be damped by the bulk viscosity of 
modified Urca process at higher temperatures and by the shear viscosity at 
lower temperatures as it is demonstrated in Fig. 9. It is to be noted here that
the onset of $K^-$ condensation is very sensitive to the rotation period
of the neutron star under investigation. The gravitational instability window 
for the case with $K^-$ condensate in nuclear matter would be different from 
the situation when hyperons are included with the condensate. 

\section{Acknowledgments}
We thank J\"urgen Schaffner-Bielich for many fruitful discussions.
DB thanks the Alexander von Humboldt Foundation for the fellowship. We also
thank Horst St\"ocker and Walter Greiner for their support and   
acknowledge the warm hospitality at the Institute for Theoretical Physics, 
J.W. Goethe University, Frankfurt am Main  and Frankfurt Institute for Advanced 
Studies (FIAS) where a part of this work was completed.

{
  
} 
\newpage
\vspace{22cm}

\begin{table}
\caption{ For neutron star matter with nucleons 
and antikaon condensate ($npK^{-}$), baryon densities $u=n/n_0$, corresponding 
to the lower ($u^l$) and upper ($u^u$) boundaries of the mixed
phase for first order $K^-$ condensation at various values of antikaon optical
potential depths $U_{\bar K}(n_0)$ (in MeV) at the saturation density 
$n_0 = 0.153$ fm$^{-3}$ are given by this table. Threshold 
densities of $\Lambda$ hyperons, $u_{\rm th}^{\Lambda}$, are shown for neutron
star matter composed of nucleons, $\Lambda$ hyperons and $K^-$ condensate. 
The values in parentheses are obtained in the presence of $\Lambda$ 
hyperons.}
\vspace {1cm}
\begin{center}
\begin{tabular}{|c|c|c|c|}

\hline
$U_{\bar K}(n_0)$ & $u^l$ & $u^u$ & $u_{\rm th}^{\Lambda}$ \\ \hline 
$$ & $$ & $$ & $$\\
$-100$ & 3.88 (4.70) & 3.88 (4.70) & 2.58 \\
$$ & $$ & $$ & $$\\
$-120$ & 3.26 (3.47) & 4.62 (5.35) & 2.58 \\
$$ & $$ & $$ & $$\\
$-140$ & 2.46 (2.46) & 5.40 (6.07) & 3.91 \\
$$ & $$ & $$ & $$\\ \hline
\end{tabular}
\end{center}
\end{table}

\vspace{2cm}

\begin{table}
\caption{Maximum masses $M_{\rm max}$, equatorial radii $R$ and their 
corresponding central baryon densities $u_{\rm cent}=n_{\rm cent}/n_0$ and 
energy densities $\varepsilon_c$ for nucleons-only ($np$) star and for 
stars with further inclusion of antikaons ($npK^-$) rotating at Kepler periods 
$P_K$ are listed below. Results at different values of antikaon optical 
potential depths $U_{\bar K}(n_0)$ (in MeV) at the saturation density are 
recorded here. The values in parentheses correspond to non-rotating neutron 
stars.}

\vspace {1cm}

\begin{center}
\begin{tabular}{|c|c|c|c|c|c|c|}

\hline
$EoS$ & $U_{\bar K}(n_0)$& $P_K (ms)$ & $u_{\rm cent}$& $\varepsilon_c (10^{15}
g/cm^3)$ & $M_{\rm max}/M_\odot$& $R (km)$ \\ \hline 
$$ & $$ & $$ & $$ & $$ & $$ & $$\\
$np$ & $-$ & 0.66 & 6.40 (7.00) & 2.15 (2.43) & 2.371 (2.019) & 14.84 (11.02)\\
$$ & $$ & $$ & $$ & $$ & $$ & $$\\
$npK^-$ & $-100$ & 0.80 & 5.33 (6.12) & 1.66 (1.96) & 2.189 (1.817) & 16.67 (11.98)\\
$$ & $$ & $$ & $$ & $$ & $$ & $$\\
$$ & $-120$ & 0.91 & 4.62 (5.20) & 1.38 (1.58) & 2.005 (1.642) & 17.73 (12.62)\\
$$ & $$ & $$ & $$ & $$ & $$ & $$\\
$$ & $-140$ & 1.06 & 3.99 (4.56) & 1.15 (1.33) & 1.648 (1.343) & 18.51 (12.98)\\
$$ & $$ & $$ & $$ & $$ & $$ & $$\\ \hline

\end{tabular}
\end{center}
\end{table}
\newpage
\vspace{-2.0cm}

{\centerline{
\epsfxsize=12cm
\epsfysize=14cm
\epsffile{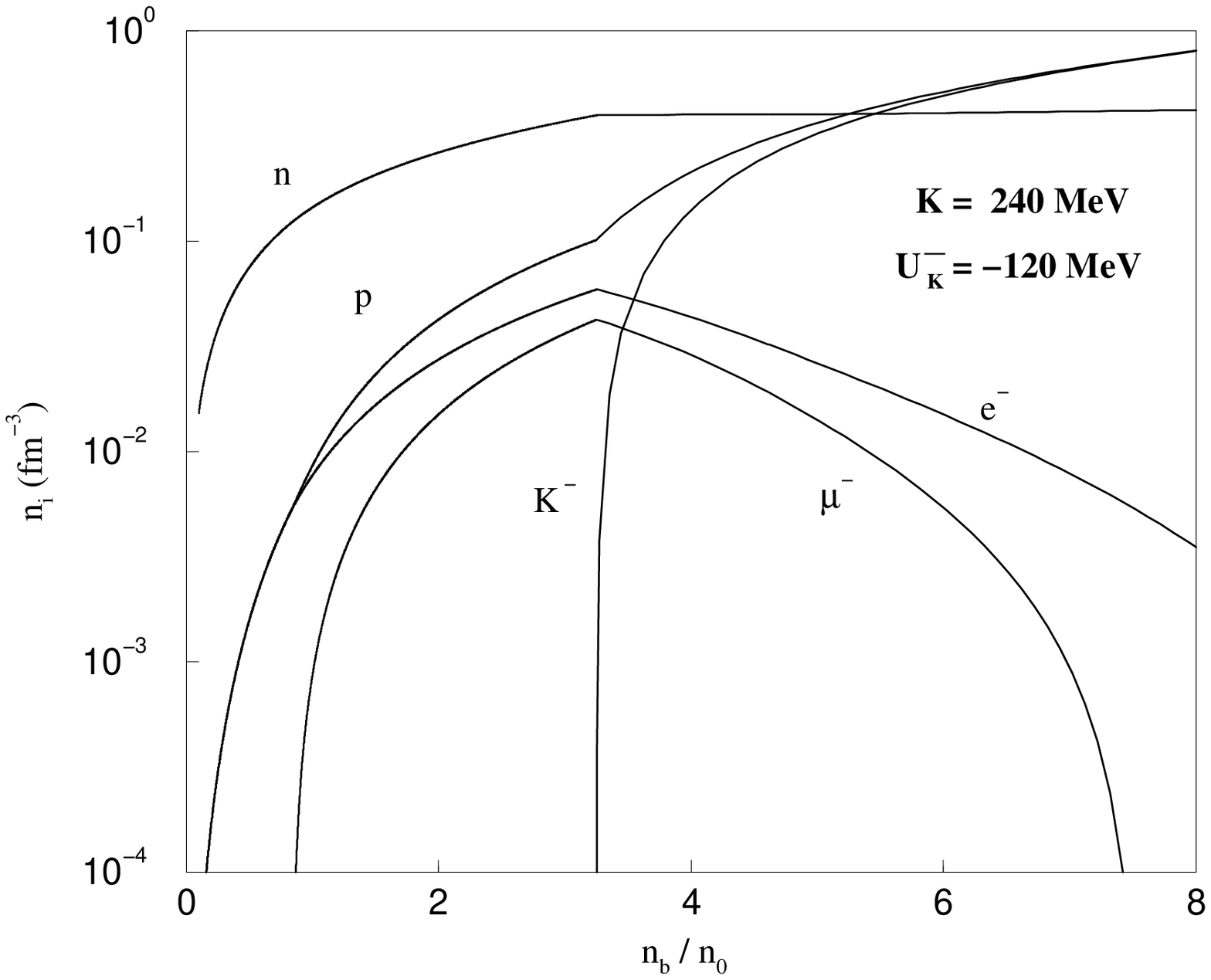}
}}

\vspace{1.0cm}

\noindent{\small{
Fig. 1. Particles abundances are plotted with normalised baryon density for 
antikaon optical potential depths at normal nuclear matter density 
$U_{\bar K}(n_0) = -120$ MeV.}}

\newpage
\vspace{-2cm}

{\centerline{
\epsfxsize=12cm
\epsfysize=14cm
\epsffile{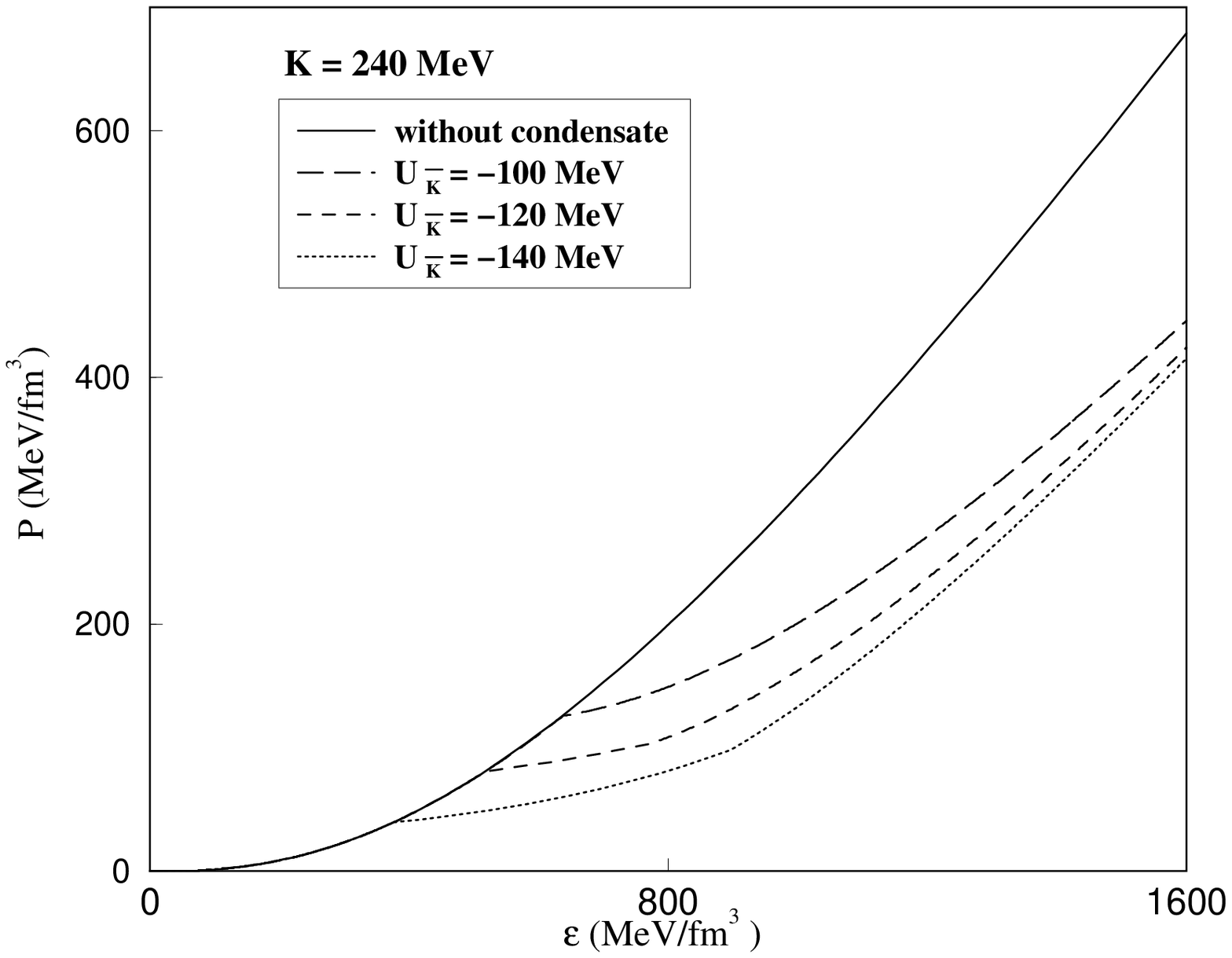}
}}

\vspace{4.0cm}

\noindent{\small{
Fig. 2. The equation of state, pressure P vs energy density $\epsilon$ , for
nucleons-only matter with and without  $K^-$ condensate. The results including
the condensate are for antikaon optical potential depths at normal nuclear 
matter density $U_{\bar K}(n_0) = -100, -120, -140$ MeV.}}

\newpage
\vspace{-2cm}

{\centerline{
\epsfxsize=12cm
\epsfysize=14cm
\epsffile{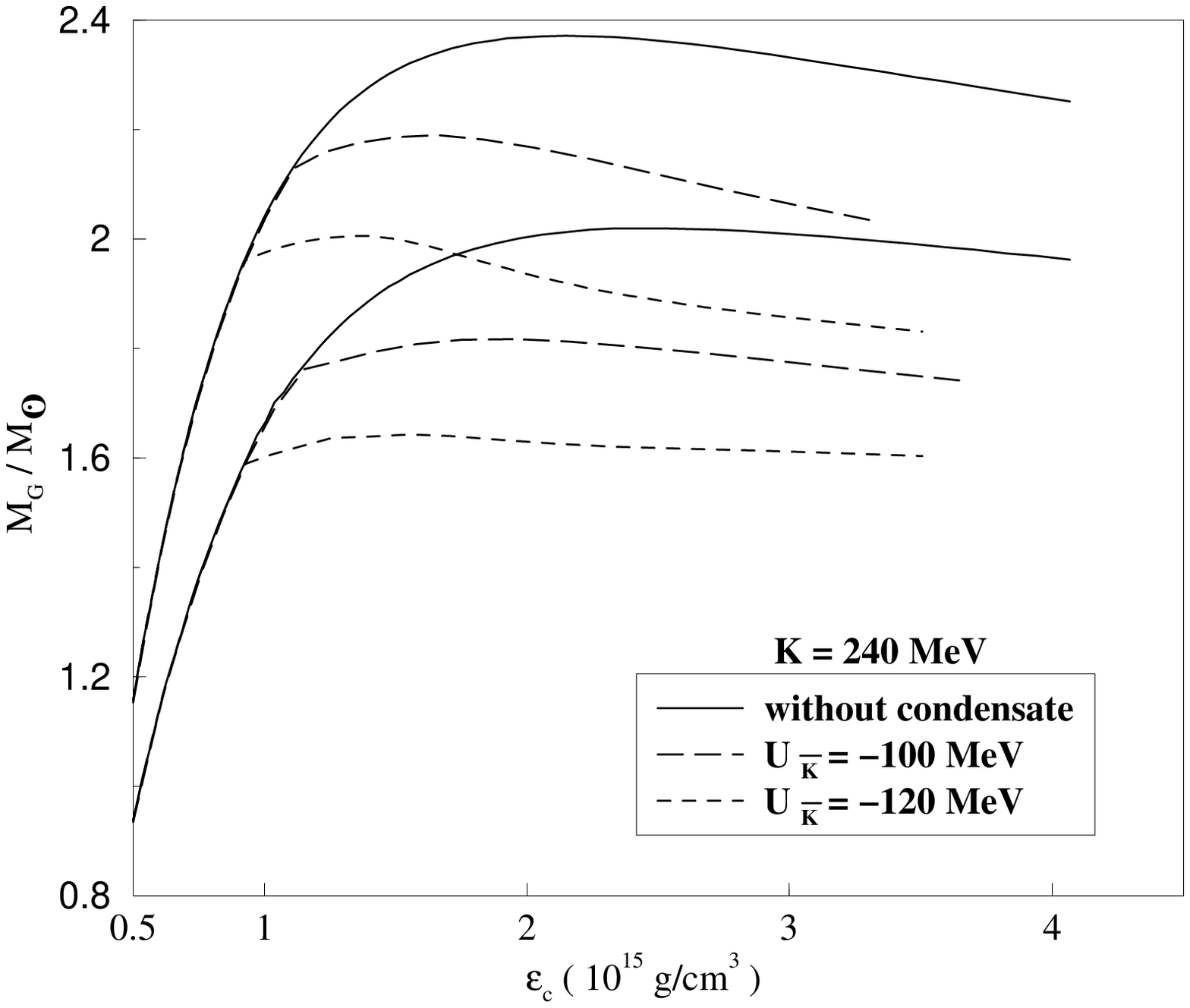}
}}

\vspace{4.0cm}

\noindent{\small{
Fig. 3. The gravitational mass for non-rotating sequence as well as mass 
shedding limit sequence of rotating neutron stars are plotted with central
energy density for the equations of state shown in Fig. 2. Different lines have
same meaning as in Fig. 2.

\newpage
\vspace{-2cm}

{\centerline{
\epsfxsize=14cm
\epsfysize=12cm
\epsffile{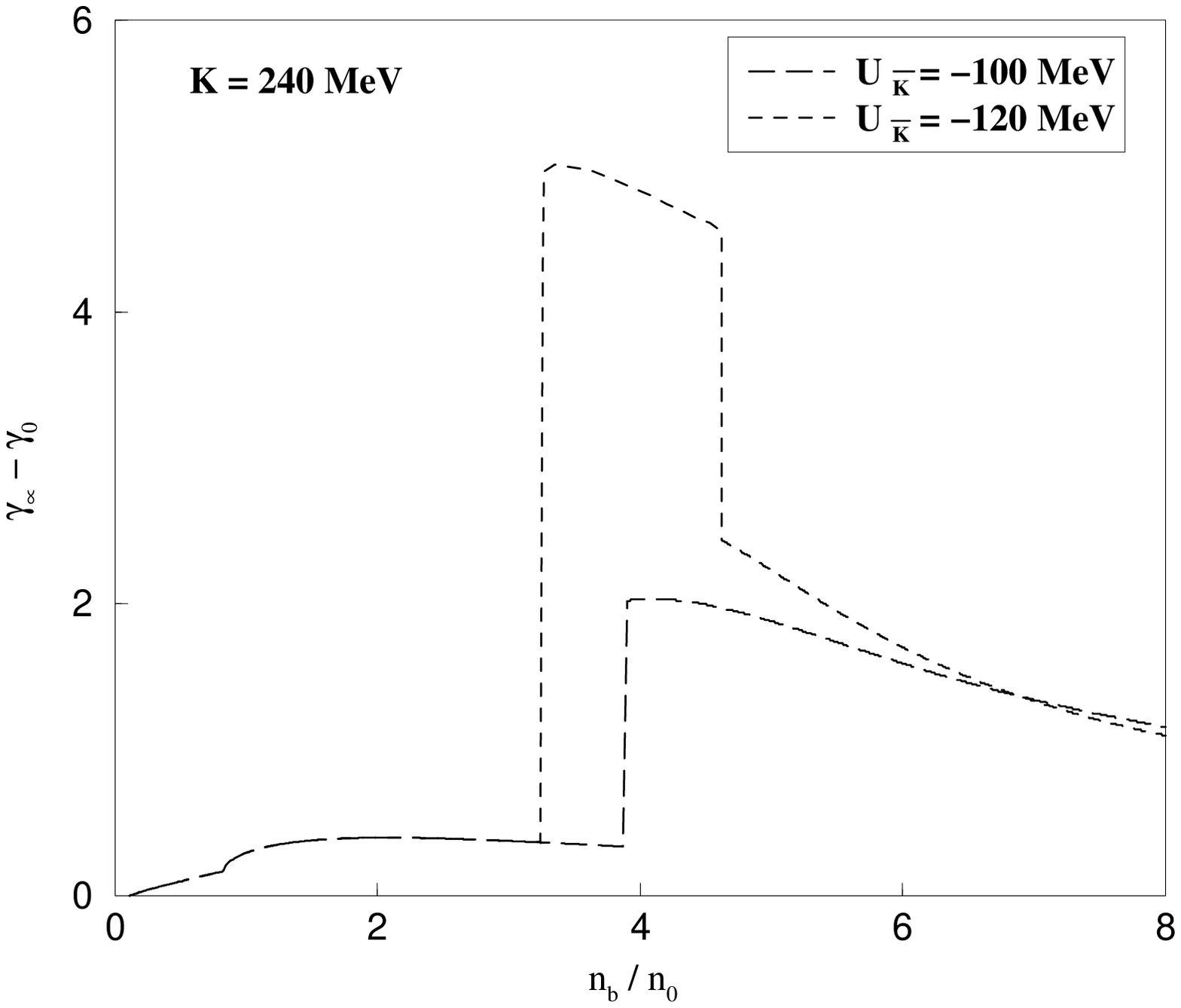}
}}

\vspace{4.0cm}

\noindent{\small{
Fig. 4. The difference of adiabatic indices, $\gamma_{\infty} - \gamma_0$, is
shown as a function of normalised baryon density for EoS including $K^-$
condensate with antikaon optical potential depths at normal nuclear 
matter density $U_{\bar K}(n_0) = -100, -120$ MeV.}}

\newpage
\vspace{-2cm}

{\centerline{
\epsfxsize=12cm
\epsfysize=14cm
\epsffile{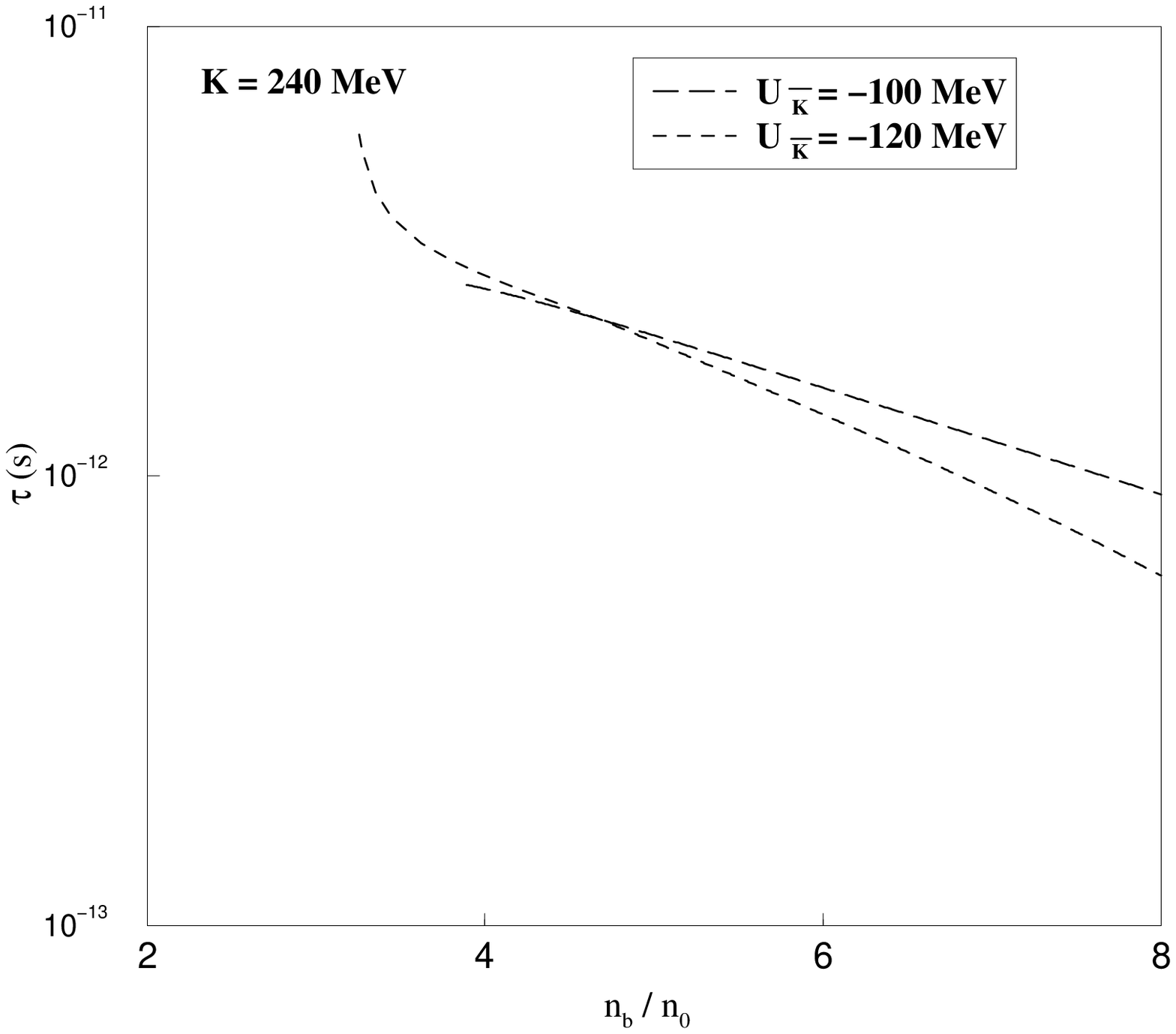}
}}

\vspace{4.0cm}

\noindent{\small{
Fig. 5. Relaxation time is plotted with normalised baryon density for the 
non-leptonic process (21) and 
antikaon optical potential depths at normal nuclear matter density 
$U_{\bar K}(n_0)= -100, -120$ MeV.}}

\newpage
\vspace{-2cm}

{\centerline{
\epsfxsize=14cm
\epsfysize=12cm
\epsffile{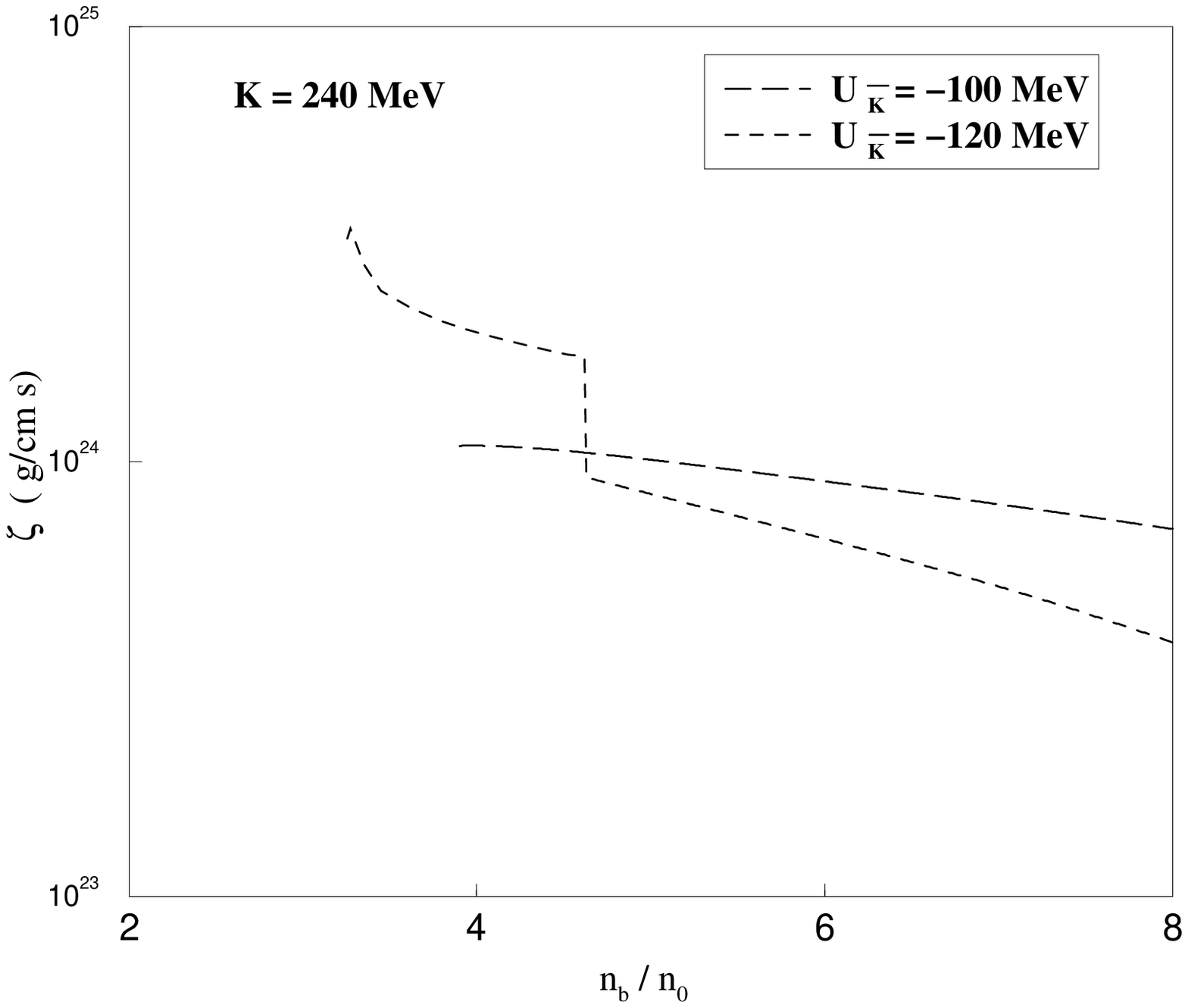}
}}

\vspace{4.0cm}

\noindent{\small{
Fig. 6. Bulk viscosity coefficient is exhibited as a function of normalised
baryon density for the process (21) and
antikaon optical potential depths at normal nuclear matter density 
$U_{\bar K}(n_0) = -100, -120$ MeV.}}

\newpage
\vspace{-2cm}

{\centerline{
\epsfxsize=14cm
\epsfysize=12cm
\epsffile{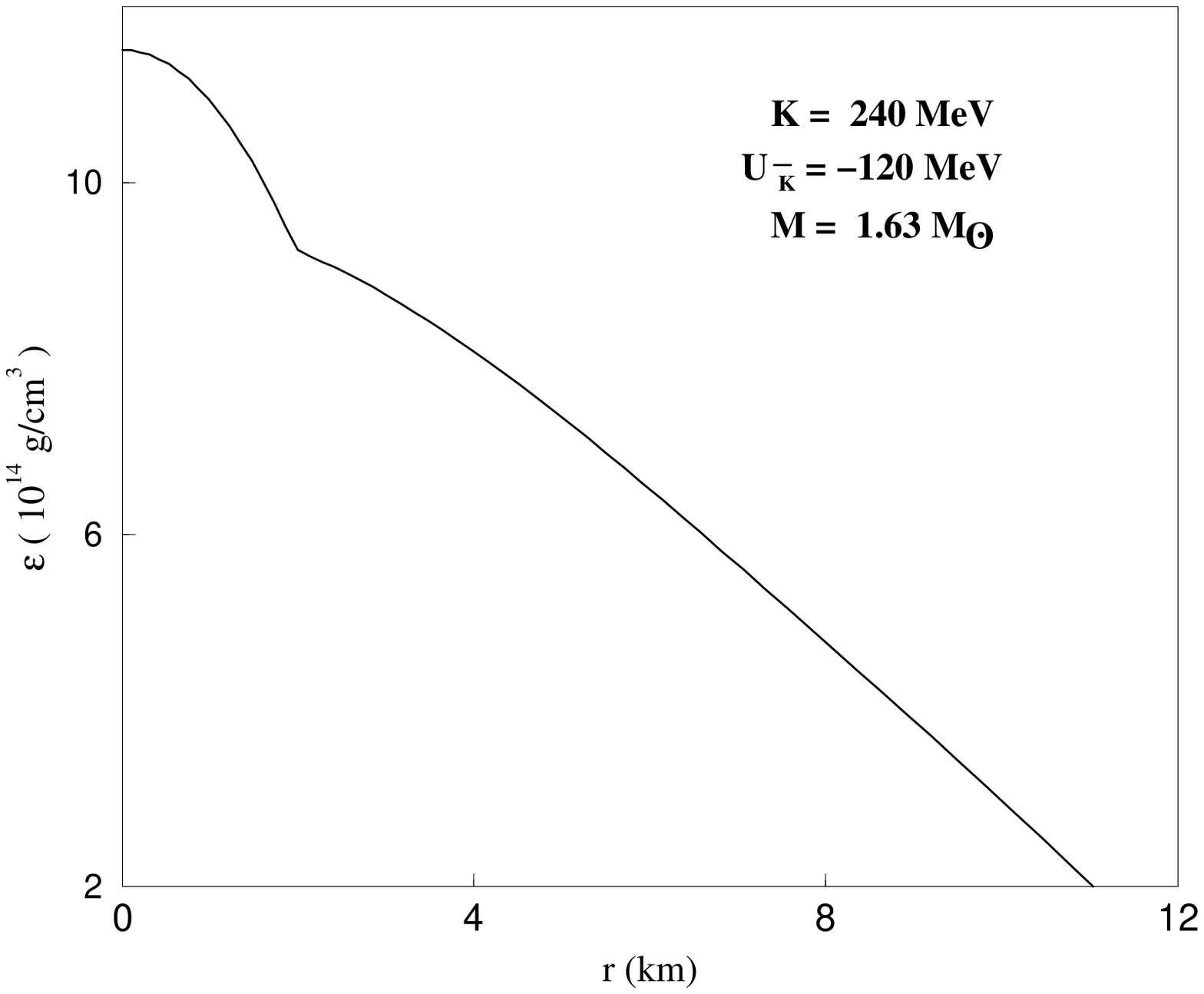}
}}

\vspace{4.0cm}

\noindent{\small{
Fig. 7. Energy density profile is shown with equatorial distance for
a rotating neutron star of mass 1.63 M$_{\odot}$ corresponding to the
EoS with antikaon optical potential depth at normal nuclear matter density 
$U_{\bar K}(n_0) = -120$ MeV.}}

\newpage
\vspace{-2cm}

{\centerline{
\epsfxsize=12cm
\epsfysize=14cm
\epsffile{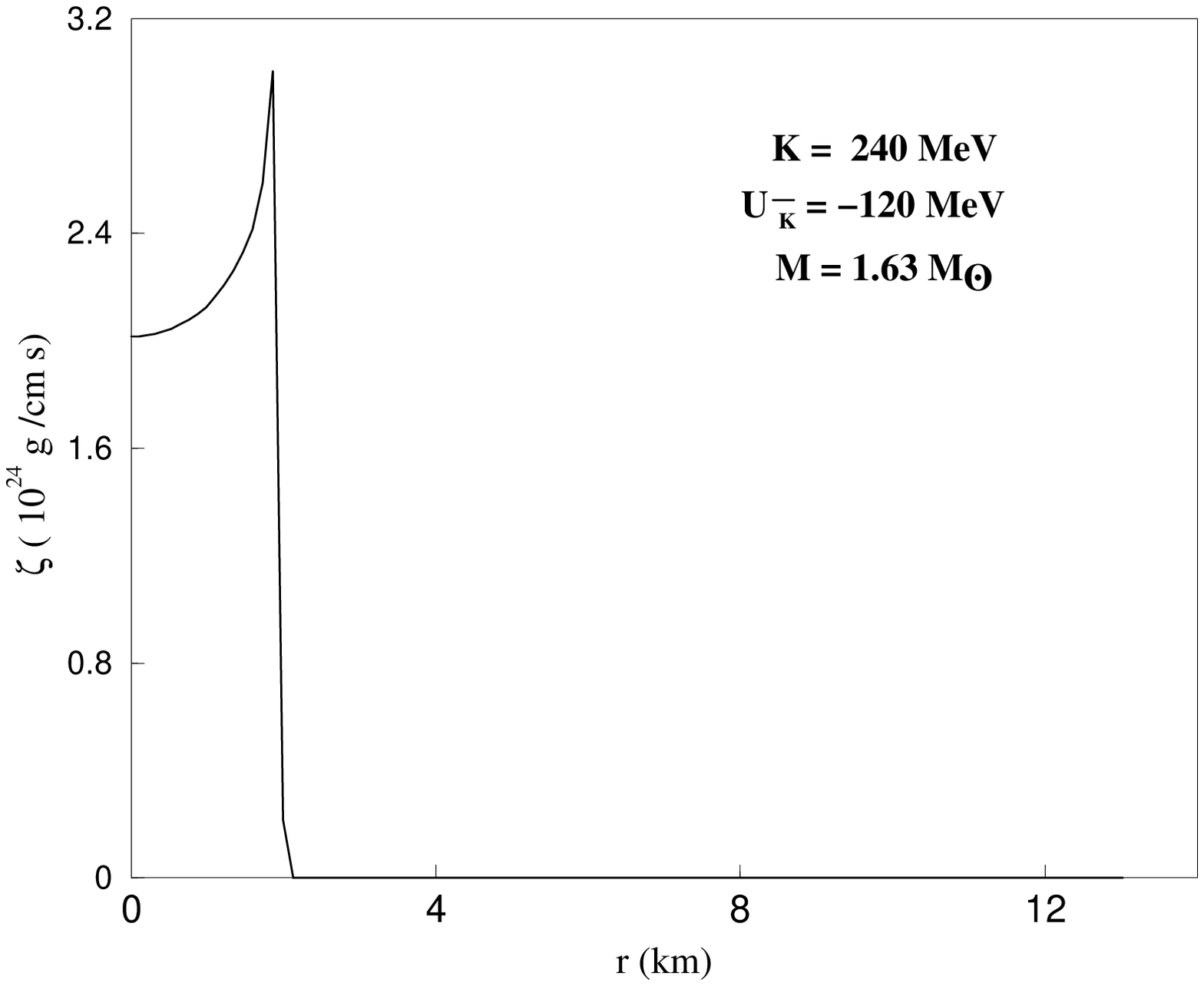}
}}

\vspace{4.0cm}

\noindent{\small{
Fig. 8. Bulk viscosity profile is shown with equatorial distance for
a rotating neutron star of mass 1.63 M$_{\odot}$. The antikaon optical 
potential depth is same as in Fig. 7.}}

\newpage
\vspace{-2cm}

{\centerline{
\epsfxsize=14cm
\epsfysize=12cm
\epsffile{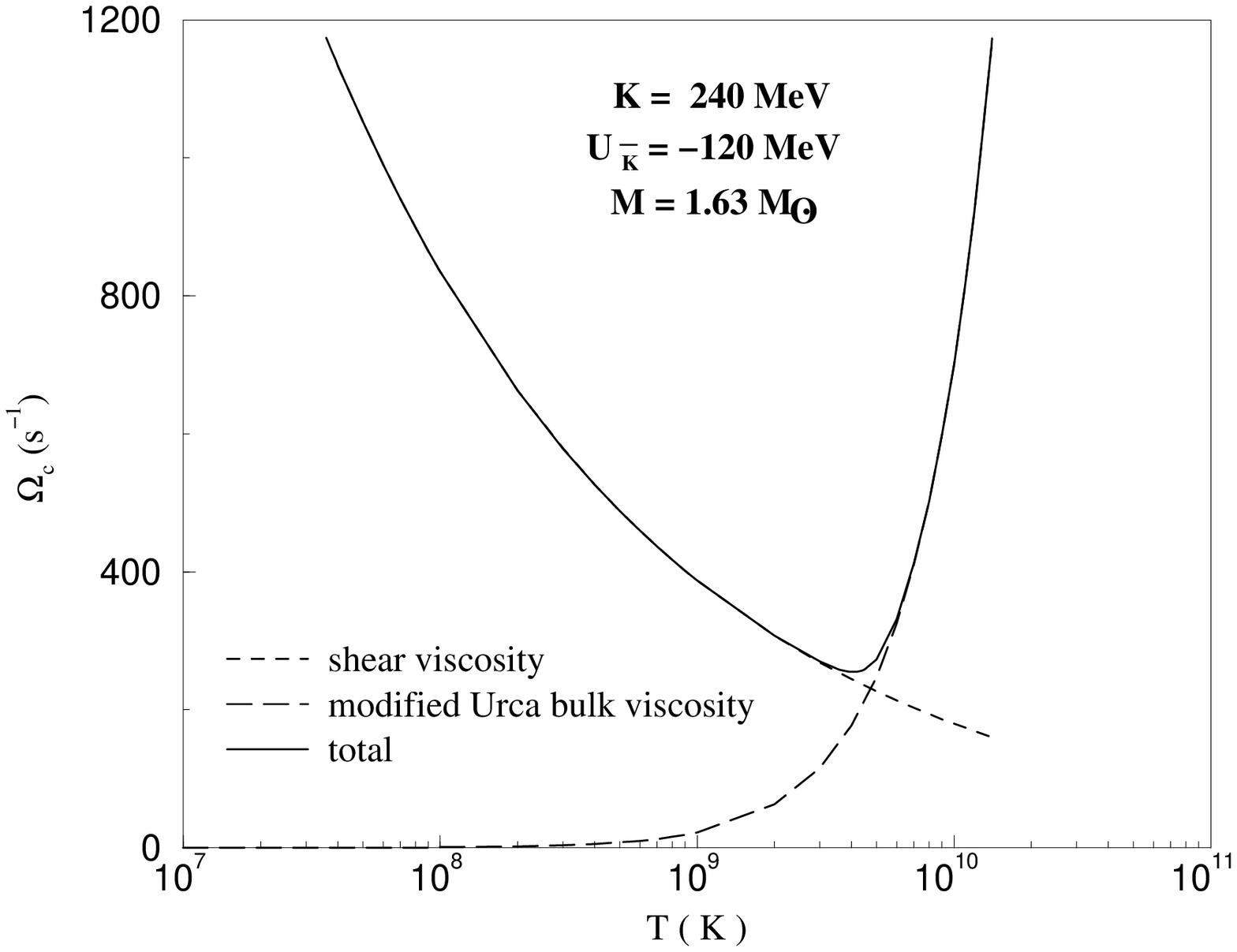}
}}

\vspace{4.0cm}

\noindent{\small{
Fig. 9. Critical angular velocity for 1.63 M$_{\odot}$ neutron star is 
plotted as a function of temperature. The antikaon optical 
potential depth is same as in Fig. 7. The solid line denotes the angular 
velocity curve when all viscous processes are included. The contributions from 
the modified Urca bulk viscosity and shear viscosity are shown by long and 
short dashed lines which merge with the solid line at higher and lower 
temperatures respectively.}}

\end{document}